\documentclass[a4paper, 10pt, conference]{IEEEtran}

\usepackage{graphics}
\usepackage{graphicx,subcaption}
\usepackage{caption}
\usepackage{epsfig,amssymb}
\usepackage{amsmath}
\usepackage{array}
\usepackage{amsmath,latexsym,amsfonts,amssymb}
\usepackage{array, tabularx}
\usepackage[table]{xcolor}
\usepackage{psfrag}
\usepackage{multirow}
\usepackage{booktabs}
\usepackage{epsfig}
\usepackage{cite}
\usepackage{tikz}
\usepackage{threeparttable}
\usepackage{url}
\usepackage{mathtools}
\usepackage{algorithm, float}
\usepackage[noend]{algpseudocode}

\newcommand{\PR}[1]{\ensuremath{\mathsf{Pr}\left\{#1\right\}}} 

\newcommand{\DC}{\ensuremath{{\rm dc}}}
\newcommand{\DP}{\ensuremath{{\rm dp}}}
\newcommand{\E}{\ensuremath{{\rm ext}}}
\newcommand{\RRH}{\ensuremath{{\rm bs}}}
\newcommand{\DHV}{\ensuremath{{\rm \V}}}
\newcommand{\DHO}{\ensuremath{{\rm \OU}}}
\newcommand{\V}{\ensuremath{{\rm ol}}}
\newcommand{\OU}{\ensuremath{{\rm ou}}}

\usepackage[labelsep=period, font=footnotesize, position=bottom, aboveskip=1pt,
belowskip=-2pt]{caption}

\DeclareMathOperator{\sinc}{sinc}
\newcommand{\Tau}{\mathrm{T}}

\newtheorem{lemma}{Lemma}

\makeatletter
\@addtoreset{case}{subsection}
\@addtoreset{condition}{theorem}
\makeatother

\IEEEoverridecommandlockouts

\begin{document}
	\title{\huge An Energy Efficient D2D Model with Guaranteed Quality of Service for Cloud Radio Access Networks}
	\author{
\IEEEauthorblockN{Isuru Janith Ranawaka\IEEEauthorrefmark{1}, Kasun T. Hemachandra\IEEEauthorrefmark{1}, Tharaka Samarasinghe\IEEEauthorrefmark{1}\IEEEauthorrefmark{2}, Theshani Nuradha\IEEEauthorrefmark{1} }
\IEEEauthorblockA{
\IEEEauthorrefmark{1}Department of Electronic and Telecommunication Engineering, University of Moratuwa, Moratuwa, Sri Lanka \\
\IEEEauthorrefmark{2}Department of Electrical and Electronic Engineering, University of Melbourne, Victoria, Australia \\
\IEEEauthorblockA{
Email: irjanith@gmail.com, kasunh@uom.lk, tharakas@uom.lk, theshanin@uom.lk
}
}
		\thanks{
This work is supported by the Senate Research Council, University of Moratuwa, Sri Lanka, under grant SRC/LT/2018/2.
 } 
	\vspace{-1.2 cm}}
	\maketitle
	
	\begin{abstract}
    This paper proposes a spectrum selection scheme and a transmit power minimization scheme for a device-to-device (D2D) network cross-laid with a cloud radio access network (CRAN). The D2D communications are allowed as an overlay to the CRAN as well as in the unlicensed industrial, scientific and medical radio (ISM) band. A link distance based scheme is proposed and closed-form approximations are derived for the link distance thresholds to select the operating band of the D2D users. Furthermore, analytical expressions are derived to calculate the minimum required transmit power to achieve a guaranteed level of quality of service in each operating band. The results demonstrate that the proposed scheme achieves nearly 50\% power saving compared to a monolithic (purely overlay or purely ISM band) D2D network.

	\end{abstract}
	
	\section{Introduction}\label{Section:Introduction}
	Cloud radio access network (CRAN) architecture has been introduced to mobile wireless networks to enable large-scale deployment and to reduce capital and operating expenditure of the network operators. However, large traffic flow in backhaul and fronthaul links can severely affect the throughput and latency performance of CRANs. To reduce the backhaul traffic in CRANs, cache enabled edge CRANs (E-CRANs) are proposed \cite{ECRAN, cluster}, while traffic offloading techniques \cite{WFIOFF1,WIFIOFF2,WIFIOFF3} are proposed to reduce the fronthaul traffic. Both these approaches require separate access points (APs) for operation, which results in additional costs for network operators. To  cater  the  demands  of  increasing user  densities, cache enabled  device-to-device (D2D)  communication  has  emerged a  promising  technology to assist the CRAN infrastructure, as a means to improve quality-of-service (QoS), throughput and energy efficiency \cite{D2D_survey2018,5GD2D_2018,HNETSOFF,D2DMm,Survey,userP_caching2019}. However, energy limitations of user devices affects the QoS of D2D networks, which motivates research on power efficient and QoS guaranteed  D2D communication protocols and user association schemes.
	
	
	
	To account for the limited energy availability at user devices, power controlling strategies have been employed to mitigate interference and provide energy efficient communication systems. In \cite{distanceBasedPowerControl2017}, a distance based power control scheme has been proposed for a D2D underlaid cellular system. A scheme to mitigate the interference generated by the D2D user equipment (UE) to the cellular UE with the help of power control of D2D UE, and also by selecting proper mode of operation based on the channel gain threshold is proposed in \cite{PK_iccW19}. 
	
	The operating spectrum band is another crucial parameter for D2D communications. Overlay, underlay and unlicensed industrial, scientific and medical radio (ISM) band D2D communications have been investigated extensively in the literature. Proper selection of spectrum band for D2D communications based on the network dynamics may further improve the QoS and energy efficiency of D2D networks.
	
	This paper proposes a spectrum selection scheme for a D2D network cross laid with an E-CRAN. The proposed scheme provides guaranteed QoS while minimizing power consumption. In contrast to a monolithic D2D network, a hybrid D2D network is proposed where D2D communications are allowed in the ISM band as well as an overlay to the E-CRAN. The contributions of this paper can be summarized as follows.
	\begin{itemize}
		\item Link distance based spectrum selection scheme is proposed for identified D2D user pairs.
		\item Link length thresholds for spectrum selection are obtained analytically.
		\item The minimum transmit power required to provide a guaranteed QoS level in each band is derived analytically.
	\end{itemize}
	The remainder of this paper is organized as follows. Section \ref{Section:System model} presents the system model under consideration and summarizes the proposed D2D communication model. Section \ref{sec:thresholdcalc} presents the link length threshold computations for each band and Section \ref{sec:txpower} gives the minimum transmit power calculation scheme for each D2D link. Numerical results obtained using the proposed scheme are shown in Section \ref{sec:results}, while Section\ref{conclusion} concludes the paper.
	\section{System Model}\label{Section:System model}
	\begin{figure}[h]
		\captionsetup{justification=centering}
		\centering
		\includegraphics[scale=0.4]{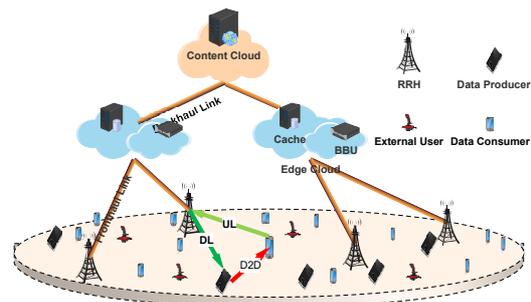}
		\caption{Communication modes and system model}
		\label{fig:model}
	\end{figure}
	We consider an E-CRAN cross laid with a  D2D network, which comprises of remote radio heads (RRHs), a content cache and a baseband unit (BBU) pool. The RRHs are spatially distributed according to a homogeneous Poisson point process (PPP) $\Phi_{\RRH}$ of intensity $\lambda_{\RRH}$. Each RRH uses a fixed transmit power $P_{\RRH}$. Three types of users, namely,  data consumers (DCs),  data producers (DPs), and external users (EUs) are considered in our model. The DCs are connected to their nearest RRH, and they request content from their connected RRH. The DPs cache the most popular content files from the edge cloud cache, such that the cache hit probability (CHP) for a given file is $p$. Moreover, we assume that a typical DC is at the origin and hereafter, we refer to it as the DC. The spatial distributions of the DCs and the DPs are modeled using homogeneous PPPs $\Phi_{\DC}$ and $\Phi_{\DP}$ with intensities of $\lambda_{\DC}$ and  $\lambda_{\DP}$, respectively. A PPP $\Phi_{\E}$ of intensity $\lambda_{\E}$ is used to model the spatial distribution of the EUs which operate in the ISM band, with a fixed power $P_{\E}$.
	
	An interference limited network is assumed where the additive noise is negligible compared to interference. For all links, Rayleigh fading is assumed where the channel power coefficients are independently and identically distributed exponential random variables (RVs) of unit mean. A distance dependent path loss model with exponent $\alpha>2$ is also used to model large-scale fading, while the effects of shadowing are neglected due to the short lengths of D2D links.
	
	We assume that all DCs will make their requests simultaneously. A request of a DC will generally be served by the RRH. However, if there is a DP in the vicinity who has the requested file in its cache, the DC may get served by this DP. Content delivery via an RRH is referred to as the cellular mode, while the delivery from a DP is referred to as the D2D mode. The D2D mode will be chosen only if it can provide equal or better QoS than the cellular mode. Since we are interested in delay sensitive content such as high definition video, the transmission delay violation probability (DVP) with respect to a given delay threshold $D_{\max}$, $\rm i.e.$, for link delay $D$, $\PR{D > D_{\max}}$, is used to measure the QoS. Intuitively, lower the DVP, higher the QoS experienced by the user.
	
	In D2D mode, the distance between the DC and the serving DP is used to determine whether the communication occur in the ISM band or as an overlay to the cellular spectrum. These two schemes are referred to as outband mode and overlay mode, respectively. In similar environments, outband DPs, who are assumed to operate at a higher carrier frequency, have a small coverage area compared to overlay DPs, who operate at a lower carrier frequency. Intuitively, the DC and DP pairs with short links are allocated to the outband mode, pairs having moderately long links are allocated to the overlay mode, and pairs with long links may not use the D2D mode as they fail to satisfy the QoS requirements. The content delivery procedure for our system model is summarized in Algorithm \ref{algo:useralloc}, where $d_{\DHO}^{\star}$ and $d_{\DHV}^{\star}$ are the distance thresholds for outband and overlay modes, respectively.
	\begin{algorithm}[h]
		\caption{Spectrum Selection  and Transmit Power Control}\label{algo:useralloc}
		\begin{algorithmic}[1]
			\For { each request 
				\State $d$ $\leftarrow$ calculate the distance between DC and DP
				\If{($d \leq d_{\DHO}^{\star}$ )}
				\State $P$ $\leftarrow$ calculate the outband power 
				\If{($ P \leq P_{\max} $)}
				\State transmit in outband network using power $P$
				\Else
				\State transmit using cellular communication
				\EndIf
				\ElsIf{($ d_{\DHO}^{\star} \leq d \leq d_{\DHV}^{\star} $ )}
				\State $P$ $\leftarrow$ calculate the overlay power 
				\If{($P \leq P_{\max} $)}
				\State transmit in overlay network using power $P$
				\Else
				\State transmit using cellular communication
				\EndIf
				\Else
				\State transmit using cellular communication
				\EndIf
				\EndFor
			}
		\end{algorithmic}
	\end{algorithm}
	
	Obtaining analytical expressions for the optimal values of $d_{\DHO}^{\star}$ and $d_{\DHV}^{\star}$, and the minimum required transmit powers of the DPs are the main contributions of this paper. The notations used in this paper are tabulated in Table \ref{tab:notation}.
	\vspace{-0.2cm}\begin{table}[ht]
		\centering
				\begin{threeparttable}
			\caption{Notation Description} \vspace{.3cm} 
			\fontsize{6}{7}\selectfont
			\label{tab:notation}
			\parbox{.2\textwidth}{
				\begin{tabular}{l c } 
					\hline\hline                        
					{\bf Description}  & {\bf Notation} \\[0.5ex]
					\hline
					Bandwidth of a cellular channel  & $B_{\RRH}$\\  
					Bandwidth of an outband channel  & $B_{\DHO}$\\
					Bandwidth of an overlay channel   & $B_{\DHV}$\\
					application level processing delay & $c$\\
					Distance from the DC to the nearest RRH & $d_{\RRH,0}$ \\
					Distance from the DC to the $k^{th}$ DP operating in outband & $d_{\DHO,k}$\\
					Distance from the DC to the $k^{th}$ DP in overlay & $d_{\DHV,k}$\\
					SIR of channel between DC to RRH  & $\gamma_{\RRH}$\\
					SIR of channel between DC to $k^{th}$ DP in outband & $\gamma_{\DHO,k}$\\
					SIR of channel between DC to $k^{th}$ DP in overlay & $\gamma_{\DHV,k}$\\
					Delay of the channel between DC and RRH & $D_{\RRH,0}$\\
					Delay of the channel between DC to the $k^{th}$ DP in outband & $D_{\DHO,k}$\\
					Delay of the channel between DC to the $k^{th}$ DP in overlay & $D_{\DHV,k}$\\
					Fading coefficient of the channel between DC and RRH & $h_{\RRH,0}$\\
					Fading coefficient of the channel between DC to the $k^{th}$ DP in outband & $h_{\DHO,k}$\\
					Fading coefficient of the channel between DC to the $k^{th}$ DP in overlay & $h_{\DHV,k}$
					
					\\[1ex]
					\hline
			\end{tabular}}
		\end{threeparttable}  
		\vspace*{-0.5cm}
	\end{table}
	\section{Spectrum Selection Scheme}\label{sec:thresholdcalc}
	The computation of distance thresholds requires several intermediate results, namely, the DVPs for each communication mode and the spatial intensities of the DPs in each D2D mode. 
	\subsection{DVP Calculation}
	We begin by considering that the DC requests a file of size $M$ from the edge cloud. The edge cloud may deliver the file directly through the RRH or via a DP. The DVPs of each mode are used to make this decision. The DVP of a link between the DC and its nearest RRH is given by the following lemma.
	\begin{lemma}
		\label{lem:DL}
		The DVP of the link between the DC and the nearest RRH is  given by \vspace*{-0.4cm}
		\begin{equation}
		\Tau_{\RRH}(D_{\max}) = \frac{(\gamma_{\RRH}^{\star})^{\frac{2}{\alpha}}}{(\gamma_{\RRH}^{\star})^{\frac{2}{\alpha}} + \sinc \left(\frac{2}{\alpha}\right)},\label{eq:undbsf}
		\end{equation}
		where $\gamma_{\RRH}^{\star} = 2^{\frac{M}{B_{\RRH}(D_{\max}-c)}}-1 $.
	\end{lemma}
	\begin{IEEEproof}
		Letting $D$ to be the sum of the transmit duration, propagation and processing delays, the DVP conditioned on $d_{\RRH,0}$, the distance between the DC and the RRH, is given by  \vspace{-0.2cm} 
		\begin{align}
		\PR{D > D_{\max}| d_{\RRH,0}} &=  \PR{\frac{M}{B_{\RRH}\log(1+\gamma_{\RRH})} +c > D_{\max}| d_{\RRH,0}}, \nonumber \\
		& = \PR{\gamma_{\RRH} < \gamma_{\RRH}^{\star}| d_{\RRH,0}}.\label{eq:gammabs}
		\end{align}
		The signal-to-interference-ratio (SIR) at the receiver for the link of interest is given by \vspace{-0.3cm}
		\begin{equation}
		\gamma_{\RRH} = \frac{P_{\RRH}h_{\RRH,0}d_{\RRH,0}^{-\alpha}}{\sum_{j \in \Phi_{\RRH}^\prime} P_{\RRH}h_{\RRH,j}d_{\RRH,j}^{-\alpha}},\label{eq:SIRBS} 
		\end{equation}
		where $\Phi_{\RRH}^\prime$ represents the point process governing the locations of the interfering RRHs. Evaluating \eqref{eq:gammabs} is well studied in the literature \cite{power}, and the DVP conditioned on $d_{\RRH,0}$ is given by 
		\begin{equation}
		\PR{D > D_{\max}\mid d_{\RRH,0}}  = 1- \exp\left({\frac{-\pi\lambda_{\RRH}(\gamma_{\RRH}^{\star})^{\frac{2}{\alpha}}d_{\RRH,0}^{2}}{\sinc \left(\frac{2}{\alpha}\right) }}\right).\label{eq:undbscond}
		\end{equation}
		Averaging (\ref{eq:undbscond}) using probability density function (PDF) of $d_{\RRH,0}$ given by $f_{d_{\RRH,0}}(r) = 2\pi\lambda_{\RRH}r\exp\left(-\pi\lambda_{\RRH}r^2\right)$, completes the proof.
	\end{IEEEproof}
		
	
	Next, $\Tau_{\RRH}(D_{\max})$ is compared with the DVP values achieved in the two D2D modes (outband and overlay). To account for worst case DVP in D2D mode, we assume that the DP containing the requested content is located at a distance equal to the threshold distance for each D2D mode. Note that (\ref{eq:undbscond}) can be used to make this comparison. However, this leads to decision thresholds which are functions of $d_{\RRH,0}$ as well. Physically, this means each DC has its own decision threshold, that depends on its distance from the RRH. This makes it prohibitively hard for us to obtain the spatial intensities of the overlay and outband DPs, $\lambda_{\DHO}$ and $\lambda_{\DHV}$, respectively, which are required to calculate the DVP values for each D2D mode. Therefore, we have averaged out the effect of $d_{\RRH,0}$ to obtain a universal distance threshold, valid for the entire network. 
	With the idea of this common threshold, next we derive $\lambda_{\DHO}$ and $\lambda_{\DHV}$. 

	To this end, we thin PPP $\Phi_{\DC}$ into three point processes to represent DCs served by RRHs, by a DP as outband and by a DP as an overlay. Moreover, we assume that outband DCs and the overlay DCs form homogeneous PPPs $\Phi_{\DHO}$ and $\Phi_{\DHV}$, respectively. Since the separation of the DCs into outband and overlay depends on the distance between DCs and DPs, the thinning of $\Phi_{\DC}$ will not result in homogeneous PPPs. However, similar approximations are used in \cite{DPPP,andrews_d2d,VAL2,VAL3} with sufficient accuracy. In Section \ref{sec:results}, we relax this assumption in simulation results, which are used to validate the analytical results. The approximate values for $\lambda_{\DHO}$ and $\lambda_{\DHV}$ are given in the following lemma.
	\begin{lemma}\label{lemma2}
		The intensities of the two point processes $\Phi_{\DHO}$ and $\Phi_{\DHV}$ are given by $\lambda_{\DHO} = \lambda_{\DC}\left[1 - e^{-p\pi\lambda_{\DP}\left(d_{\DHO}^{\star}\right)^{2}}\right]$ and $\lambda_{\DHV} = \lambda_{\DC}\left[e^{-p\pi\lambda_{\DP}\left(d_{\DHO}^{\star}\right)^{2}}-e^{-p\pi\lambda_{\DP}\left(d_{\DHV}^{\star}\right)^{2}}\right]$, respectively.
	\end{lemma}
	\begin{IEEEproof}
		Using the null probability of $\Phi_{\DP}$, the probability of existence of a DP containing the requested file within the distance of $d_{\DHO}^{\star}$ from the DC is given by $\left[1 - e^{-p\pi\lambda_{\DP}(d_{\DHO}^{\star})^{2}}\right]$, where we have assumed the intensity of the DPs containing the requested file is $p \lambda_{\DP}$. The edge cloud randomly selects a DP within the distance of $d_{\DHO}^{\star}$ from the DC, which will transmit in outband. Hence, multiplying the probability by $\lambda_{\DC}$ gives the intensity of DPs, who are eligible to transmit in outband. Assuming a one to one mapping of DCs to DPs, this intensity is equal to $\lambda_{\DHO}$.
		
		Similarly, the probability of existence of a DP having the requested content between the distance of $d_{\DHO}^{\star}$ and $d_{\DHV}^{\star}$ is given by $\left[1 - e^{-p\pi\lambda_{\DP}(d_{\DHV}^{\star})^{2}}\right]-\left[1 - e^{-p\pi\lambda_{\DP}(d_{\DHO}^{\star})^{2}}\right]$. The randomly selected DP will transmit to the DC in the overlay band. Therefore, multiplying this probability by $\lambda_{\DC}$ gives $\lambda_{\DHV}$.
	\end{IEEEproof}

	Using the approximate intensities $\lambda_{\DHO}$ and $\lambda_{\DHV}$, Lemma \ref{lem:DL2} gives the conditional DVPs (conditional on D2D link lengths) of typical D2D links in outband and overlay modes.
	\begin{lemma}\label{lem:DL2}
		The DVPs of a typical outband D2D link ($k^{\textnormal{th}}$), and a typical overlay D2D link are  given by $ \PR{D_{\DHO,k} > D_{\max}\mid d_{\DHO,k}}  = \Tau_{\DHO,k}(D_{\max},d_{\DHO,k})$ and $ \PR{D_{\DHV,k} > D_{\max}\mid d_{\DHV,k}}  = \Tau_{\DHV,k}(D_{\max},d_{\DHV,k})$, respectively, where 	\vspace{-0.3cm}
		\begin{multline}
		\Tau_{\DHO,k}(D_{\max},d_{\DHO,k}) \\ = 1- \exp\left(\frac{-\pi\left[\lambda_{\DHO}E\left(P_{\DHO,j}^{\frac{2}{\alpha}}\right)+P_{\E}^{\frac{2}{\alpha}}\lambda_{\E}\right]\left(\gamma_{\DHO}^{\star}\right)^{\frac{2}{\alpha}}d_{\DHO,k}^{2}}{\sinc \left(\frac{2}{\alpha}\right)P_{\DHO,k}^{\frac{2}{\alpha}}}\right)
		\end{multline}
		\begin{equation}
		\Tau_{\DHV,k}\left(D_{\max},d_{\DHV,k}\right) = 1- \exp\left({\frac{-\pi\lambda_{\DHV}E\left(P_{\DHV,j}^{\frac{2}{\alpha}}\right)(\gamma_{\DHV}^{\star})^{\frac{2}{\alpha}}d_{\DHV,k}^{2}}{\sinc \left(\frac{2}{\alpha}\right)P_{\DHV,k}^{\frac{2}{\alpha}}}}\right).
		\end{equation}
	\end{lemma}
	\begin{IEEEproof}
		Following the  proof of Lemma \ref{lem:DL}, we have
		\begin{align}
		\Tau_{\DHO,k}\left(D_{\max},d_{\DHO,k}\right) &= \PR{\gamma_{\DHO,k} < \gamma_{\DHO}^{\star}\mid d_{\DHO,k}},\label{eq:DVPOU}\\
		\Tau_{\DHV,k}\left(D_{\max},d_{\DHV,k}\right) &= \PR{\gamma_{\DHV,k} < \gamma_{\DHV}^{\star}\mid d_{\DHV,k}}\label{eq:DVPOL},
		\end{align}
		where $\gamma_{\DHO}^{\star} = 2^{\frac{M}{B_{\DHO,k}(D_{\max}-c)}}-1$ 
		and $\gamma_{\DHV}^{\star} = 2^{\frac{M}{B_{\DHV,k}(D_{\max}-c)}}-1 $. 
		The SIRs of D2D links in each mode can be given as
		\begin{align}
		\gamma_{\DHO,k} &= \frac{P_{\DHO,k}h_{\DHO,k}d_{\DHO,k}^{-\alpha}}{\sum_{j \in \Phi_{\DHO}^{\prime} } P_{\DHO,j}h_{\DHO,j}d_{\DHO,j}^{-\alpha} + \sum_{j \in \Phi_{\E} } P_{\E,j}h_{\E,j}d_{\E,j}^{-\alpha}}, \label{eq:SIROU}\\
		\gamma_{\DHV,k} &= \frac{P_{\DHV,k}h_{\DHV,k}d_{\DHV,k}^{-\alpha}}{\sum_{j \in \Phi_{\DHV}^{\prime} } P_{\DHV,j}h_{\DHV,j}d_{\DHV,j}^{-\alpha}}, \label{eq:SIROL}
		\end{align}
		where $\Phi_{\DHO}^{\prime}$  and $\Phi_{\DHV}^{\prime}$ are the PPPs governing the locations of the interfering 
		DPs in $\Phi_{\DHO}$ and $\Phi_{\DHV}$, respectively. Evaluating  \eqref{eq:DVPOU} and \eqref{eq:DVPOL} using \eqref{eq:SIROU} and  \eqref{eq:SIROL} as in the proof of Lemma \ref{lem:DL} concludes the proof.
	\end{IEEEproof}
\subsection{Link Length Threshold Calculation}
	Using the DP intensities and DVPs of each D2D mode, expressions for the link length thresholds  $d_{\DHO}^{\star}$ and $d_{\DHV}^{\star}$ can be found as shown in Lemma \ref{lem:DistTh}.	\begin{lemma}\label{lem:DistTh} 
		The outband and overlay distance thresholds are given by
		$d_{\DHO}^{\star} = \left(\frac{\sqrt{B^2+4AC} -B}{2A}\right)^{\frac{1}{2}}$, $d_{\DHV}^{\star} = \left(\frac{\sqrt{E^2+4DC} -E}{2D}\right)^{\frac{1}{2}},$
		where $A = \frac{\pi^2 p(\gamma_{\DHO}^{\star})^{\frac{2}{\alpha}}\lambda_{\DP}\lambda_{\DC}E\left(P_{\DHO,j}^{\frac{2}{\alpha}}\right)}{\sinc \left(\frac{2}{\alpha}\right)P_{\max}^{\frac{2}{\alpha}}}$, $B = \frac{\pi\left(\gamma_{\DHO}^{\star}\right)^{\frac{2}{\alpha}}P_{\E}^{\frac{2}{\alpha}}\lambda_{\E}}{\sinc\left( \frac{2}{\alpha} \right)P_{\max}^{\frac{2}{\alpha}}}$, $C = \mid\ln\left(1-\Tau_{\RRH}\left(D_{\max}\right)\right)\mid$, $D = \frac{\pi^{2}p\lambda_{\DC}\lambda_{\DP}E\left(P_{\DHV,j}^{\frac{2}{\alpha}}\right)\left(\gamma_{\DHV}^{\star}\right)^{\frac{2}{\alpha}}}{\sinc \left( \frac{2}{\alpha}\right)P_{\max}^{\frac{2}{\alpha}}}$ and $E =\frac{\pi p\lambda_{\DC}E\left(P_{\DHV,j}^{\frac{2}{\alpha}}\right)\left(\gamma_{\DHV}^{\star}\right)^{\frac{2}{\alpha}}\left[e^{-\pi\lambda_{\DP}\left(d_{\DHO}^{\star}\right)^{2}}-1\right]}{\sinc \left(\frac{2}{\alpha}\right)P_{\max}^{\frac{2}\alpha}}$.
	\end{lemma}
	
	\begin{IEEEproof}
		We assume that the selected DP is at the distance threshold $d_{\DHO}^{\star}$ from the DC. To be eligible for a viable outband D2D link while using the maximum available transmit power $P_{\max}$, this DP should be able to satisfy the QoS requirement
		\begin{align}\label{qos_outband}
		\Tau_{\DHO,k}\left(D_{\max},d_{\DHO}^{\star}\right) \leq T_{\RRH}(D_{\max}). 
		\end{align}
		By substituting the results of Lemma \ref{lemma2} and \ref{lem:DL2}, and by applying the first order Taylor series approximation  $e^{-ax} = 1-ax$, \eqref{qos_outband} can be simplified as
		\begin{multline}
		\frac{\pi\left[\pi p \lambda_{\DP}\lambda_{\DC}(d_{\DHO}^{\star})^{2}E\left(P_{\DHO,j}^{\frac{2}{\alpha}}\right)+P_{\E}^{\frac{2}{\alpha}}\lambda_{\E}\right](\gamma_{\DHO}^{\star})^{\frac{2}{\alpha}}(d_{\DHO}^{\star})^{2}}{\sinc \left(\frac{2}{\alpha}\right)P_{\max}^{\frac{2}{\alpha}}}  \\ \leq \mid\ln\left(1-\Tau_{\RRH}\left(D_{\max}\right)\right)\mid.\label{eq:prro2}
		\end{multline}
		Solving \eqref{eq:prro2}, gives us $d_{\DHO}^{\star}$. Same procedure can be used to obtain an expression for $d_{\DHV}^{\star}$.
	\end{IEEEproof}

	Clearly, $d_{\DHO}^{\star}$ and $d_{\DHV}^{\star}$ depend on $E\left(P_{\DHO,j}^{\frac{2}{\alpha}}\right)$ and $E\left(P_{\DHV,j}^{\frac{2}{\alpha}}\right)$, which depend on the transmit powers of the other DPs in each band. Obtaining analytical expressions for these expectations appears to be intractable since the PDF of the transmit powers of the DPs is not known. Therefore, assuming worst case conditions, the interferes are allowed to transmit at their maximum power, making  $E\left(P_{\DHO,j}^{\frac{2}{\alpha}}\right) = P_{\max}^{\frac{2}{\alpha}}$ and $E\left(P_{\DHV,j}^{\frac{2}{\alpha}}\right) = P_{\max}^{\frac{2}{\alpha}}$. This simplifies $d_{\DHO}^{\star}$ and $d_{\DHV}^{\star}$ such that $A = \frac{\pi^{2} p(\gamma_{\DHO}^{\star})^{\frac{2}{\alpha}}\lambda_{\DP}\lambda_{\DC}}{\sinc \left(\frac{2}{\alpha}\right)}$ ,$D = \frac{\pi^{2} p(\gamma_{\DHV}^{\star})^{\frac{2}{\alpha}}\lambda_{\DP}\lambda_{\DC}}{\sinc \left(\frac{2}{\alpha}\right)}$, and $E =\frac{\pi p\lambda_{\DC}\left(\gamma_{\DHV}^{\star}\right)^{\frac{2}{\alpha}}\left[e^{-\pi\lambda_{\DP}\left(d_{\DHO}^{\star}\right)^{2}}-1\right]}{\sinc \left(\frac{2}{\alpha}\right)}$, which result in lower bounds for $d_{\DHO}^{\star}$ and $d_{\DHV}^{\star}$. 
	
	The thresholds can be further refined in a system setting by using an iterative computation scheme. Initially, the distance thresholds and the transmit power of each DP are calculated under the worst case conditions. In the next iteration, $E\left(P_{\DHO,j}^{\frac{2}{\alpha}}\right)$ and $E\left(P_{\DHV,j}^{\frac{2}{\alpha}}\right)$ are evaluated using the transmit powers of the previous iteration, and the distance thresholds and the transmit power of each DP are recalculated. This procedure is repeated until the distance thresholds converge to a fixed value. We refer to this approach as ``iterative optimization" in our numerical results.
	
	From the distance threshold expressions, one can deduce that when ${d_{\DHO}^{\star} \rightarrow 0}$, all DPs will be allocated to overlay band. Since reducing the outband threshold will allocate more DPs into the overlay network, the interference in the overlay band will increase. Therefore, when ${d_{\DHO}^{\star} \rightarrow 0}$, $d_{\DHV}^{\star}$ also decays exponentially. Furthermore, when $d_{\DHO}^{\star}$ increases, $d_{\DHV}^{\star}$ also increases. When the outband region expands, more users are allocated to the outband. Hence, the interference in the overlay region will be reduced, providing more communication opportunities in the overlay band.
	\section{Transmit Power Computation}\label{sec:txpower}
	Next, we calculate the parameter $P$ in Algorithm \ref{algo:useralloc}, which is the required minimum transmit power of each DP to satisfy the QoS requirement. Assume that the $k^{\textnormal{th}}$ DP containing the requested file is selected to serve the DC. We first decide on the operating band of the DP by comparing the link length with the distance thresholds. Next, the required minimum power of each DP is computed such that the DVP with a D2D link is at most equal to the DVP of delivering content through an RRH. 
	
	The following lemma formally states the required minimum power for a selected DP in each band to achieve a DVP equal to the cellular mode.
	\begin{lemma}\label{lem:powerCalc} 
		The minimum transmit power of the $k^{th}$ DP allocated to the outband network or the overlay network can be given as
		\begin{subequations}
			\begin{equation}
			\label{outband_opt}
			P_{\DHO,k}^{\prime} = \left[\frac{{\lambda_{\DHO}E\left(P_{\DHO,j}^{\frac{2}{\alpha}}\right)+P_{\E}^{\frac{2}{\alpha}}\lambda_{\E}}}{\lambda_{\RRH}}\right]^{\frac{\alpha}{2}}\left(\frac{\gamma_{\DHO}^{\star}}{\gamma_{\RRH}^{\star}}\right)\left(\frac{d_{\DHO,k}}{d_{\RRH,0}}\right)^{\alpha}
			\end{equation}
			\begin{equation}\label{overlay_opt}
			P_{\DHV,k}^{\prime} = \left[\frac{\lambda_{\DHV}E\left(P_{\DHV,j}^{\frac{2}{\alpha}}\right)}{\lambda_{\RRH}}\right]^{\frac{\alpha}{2}}\left(\frac{\gamma_{\DHV}^{\star}}{\gamma_{\RRH}^{\star}}\right)\left(\frac{d_{\DHV,k}}{d_{\RRH,0}}\right)^{\alpha}.
			\end{equation}
		\end{subequations}
	\end{lemma}
	\begin{IEEEproof}
		We first consider an outband DP. To achieve equal or better QoS compared to the cellular mode, we need
		\begin{align}
		& T_{\DHO,k}\left(D_{\max},d_{\DHO,k}\right) \leq T_{\RRH}\left(D_{\max},d_{\RRH,0}\right).\nonumber \end{align}
		By substituting the DVP values, we have
		\begin{multline}
		 \exp\left(\frac{-\pi\left[\lambda_{\DHO}E\left(P_{\DHO,j}^{\frac{2}{\alpha}}\right)+P_{\E}^{\frac{2}{\alpha}}\lambda_{\E}\right]\left(\gamma_{\DHO}^{\star}\right)^{\frac{2}{\alpha}}d_{\DHO,k}^{2}}{\sinc \left(\frac{2}{\alpha}\right)P_{\DHO,k}^{\frac{2}{\alpha}}}\right) \geq \\
		\exp\left({\frac{-\pi\lambda_{\RRH}(\gamma_{\RRH}^{\star})^{\frac{2}{\alpha}}d_{\RRH,0}^{2}}{\sinc 
				\left(\frac{2}{\alpha}\right)}}\right). \label{eq:oupw}
		\end{multline}
		Solving \eqref{eq:oupw} for $P_{\DHO,k}$ yields \eqref{outband_opt} as the minimum required transmit power for the DP. A similar approach can be used to obtain \eqref{overlay_opt}.
	\end{IEEEproof}
	One can observe that $P_{\DHO,k}$ and $P_{\DHV,k}$ depend on the ratio $\frac{d_{\DHO,k}}{d_{\RRH,0}}$ and the mean transmit power of the DPs in the operating band. 
	
	The distance thresholds identify the feasible set of outband DPs and the overlay DPs. However, since the threshold values are based on the average DVP of the cellular mode (averaged over the distance to the nearest RRH), all DPs in the feasible set may not be able to satisfy the maximum transmit power constraint for individual links. Therefore, the DPs in feasible set are individually checked for maximum power constraint violation. The DCs with selected DPs who are not capable of satisfying the power constraint will be re-allocated to the cellular mode. This refines the DP intensities in each band. The refined DP intensities in each band are given in the following lemma.
	\begin{lemma}\label{lem:EI}
		Refined intensities of the outband and the overlay band are given by
		\begin{equation}
		\label{refined_out}
		\lambda_{\DHO}^{th} = \left[\frac{12\lambda_{\DHO}}{\pi^{3}\lambda_{\RRH}^{3}\left(d_{\DHO}^{\star}\right)^{2}}\right]\left(\frac{P_{\max}}{\beta}\right)^{\frac{2}{\alpha}},
		\end{equation}
		\begin{equation}
		\label{refined_over}
		\lambda_{\DHV}^{th} = \frac{\lambda_{\DHV}}{\left(\left(d_{\DHV}^{\star}\right)^{2} - \left(d_{\DHO}^{\star}\right)^{2}\right)}\left[\frac{12}{\pi^{3}\lambda_{\RRH}^{3}}\left(\frac{P_{\max}}{\beta}\right)^{\frac{2}{\alpha}}-\left(d_{\DHO}^{\star}\right)^{2}\right],
		\end{equation}  
		where $\beta = \left[\frac{\left(\lambda_{\DHO}P_{\max}^{\frac{2}{\alpha}}+P_{\E}^{\frac{2}{\alpha}}\lambda_{\E}\right)}{\lambda_{\RRH}}\right]^{\frac{\alpha}{2}}\left(\frac{\gamma_{\DHO}^{\star}}{\gamma_{\RRH}^{\star}}\right)$ and $\eta = \left[\frac{\lambda_{\DHV}P_{\max}^{\frac{2}{\alpha}}}{\lambda_{\RRH}}\right]^{\frac{\alpha}{2}}\left(\frac{\gamma_{\DHV}^{\star}}{\gamma_{\RRH}^{\star}}\right)$. 
	\end{lemma}
	
	\begin{IEEEproof}
		The refined intensity of the outband DPs can be found as,
		\begin{align}
		\lambda_{\DHO}^{th} &= \int_{0}^{\infty} \lambda_{\DHO}\PR{P_{\DHO,k}\leq P_{\max}\mid r}f_{d_{\RRH,0}}(r)dr \nonumber  \\
		&= \int_{0}^{\infty} \lambda_{\DHO}\PR{d_{\DHO,k} \leq \left({\frac{P_{\max}}{\beta}}\right)^{\frac{1}{\alpha}}r}f_{d_{\RRH,0}}(r)dr \nonumber  \\
		&\stackrel{(a)}{=} \int_{0}^{\infty} \lambda_{\DHO}
		\left[\frac{\left(\frac{P_{\max}}{\beta}\right)^{\frac{2}{\alpha}}r^{2}}{\left(d_{\DHO}^{\star}\right)^{2}}\right]2\pi\lambda_{\RRH}re^{-\pi\lambda_{\RRH}r^2}dr. \label{eq:inEI}
		\end{align}
		where $(a)$ follows from $\PR{P_{\DHO,k}\leq x}=\frac{x^2}{\left(d_{\DHO}^{\star}\right)^{2}}$.
		Evaluating \eqref{eq:inEI}, yields \eqref{refined_out}. By following a similar approach and by using $\PR{d_{\DHV,k} \leq r} = \frac{r^{2}-\left(d_{\DHO}^{\star}\right)^{2}}{\left(d_{\DHV}^{\star}\right)^{2} -\left(d_{\DHO}^{\star}\right)^{2}}$, one can obtain \eqref{refined_over}. We have assumed the maximum interference in each band when calculating the refined intensities. 
	\end{IEEEproof}	
	The refined intensities can be used to evaluate other performance metrics such as coverage probability, average achievable rate and transmission capacity of the network. However, due to page length restrictions, we do not include those results in this version.
	\vspace*{-0.1cm}
	\section{Numerical  Results}\label{sec:results}
	In this section, we provide numerical and simulation results to validate our assumptions and to identify the benefits of the proposed algorithm. Note that the assumptions made for analysis are relaxed in simulation results. The parameters used in the simulations are tabulated in Table II.
	\vspace{-0.1cm}\begin{table}[ht]
		\centering
		\begin{threeparttable}
			\label{tab:param}
			\caption{Simulation Parameters} 
			\fontsize{7}{8}\selectfont
			\parbox{.2\textwidth}{
				\begin{tabular}{c c } 
					\hline\hline                        
					Parameter  & Value \\[0.5ex]
					\hline
					RRH power $(P_{\RRH})$ & $100mW$\\              
					Maximum power of an end device $(P_{\max})$ & $2.5mW$\\
					Power of an external user $(P_{\E})$ & $ 2mW$ \\
					Radius of the simulated area $(R)$  & $3000m$\\
					DP intensity$(\lambda_{\DP})$ & $10^{-4}$\\
					DC intensity$(\lambda_{\DC})$ & $10^{-3}$\\
					EU intensity $(\lambda_{\E})$ & $10^{-3.5}$\\
					RRH intensity $(\lambda_{\RRH})$ & $10^{-5.5}$\\
					Path loss exponent $(\alpha)$  & $3.5$\\
					File size $(M)$ & $80kB$\\
					Channel bandwith $(B_{\RRH}, B_{\DHO}, B_{\DHV})$ & $5MHz$\\
					Application level delay threshold $(D_{\max})$  & $0.5ms$\\
					Processing delay $(c)$ & $0.1ms$\\
					
					\hline
			\end{tabular}}
		\end{threeparttable}   
	\end{table}
	\begin{figure}[t]
		\centering{\includegraphics[width=0.75\linewidth]{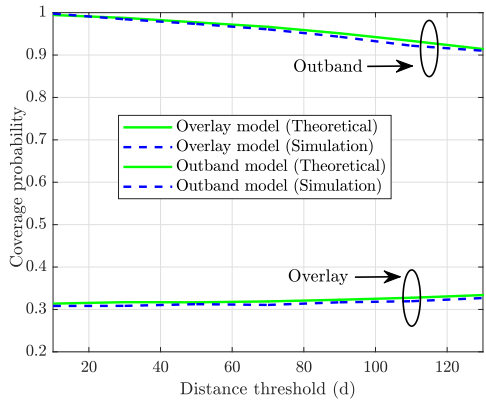}}\caption{Validation of the independent thinning approximation} \label{Fig: modelval}
	\end{figure}
	\begin{figure}[t]
	\vspace{-0.2cm}
		\centering{\includegraphics[width=0.75\linewidth]{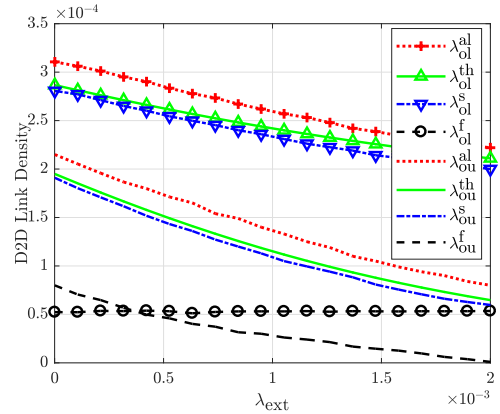}}\caption{Intensity of each D2D network against the external user intensity}\vspace{-0.3cm} \label{Fig: inten_ext}
	\end{figure}
	\begin{figure}[t]
		\centering{\includegraphics[width=0.75\linewidth]{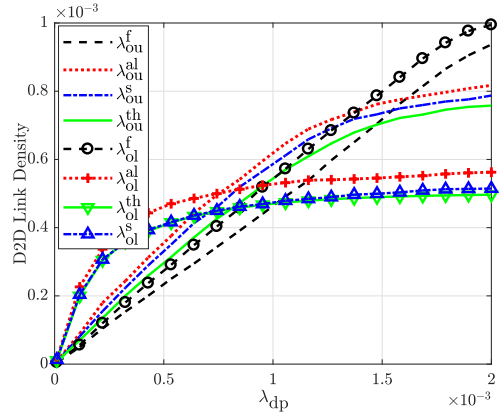}} \caption{Intensity of each D2D network against the DP intensity}\vspace{-0.3cm}\label{Fig: inten_dp}
	\end{figure}
	\begin{figure}[t]
	\vspace{-0.2cm}
		\centering{\includegraphics[width=0.75\linewidth]{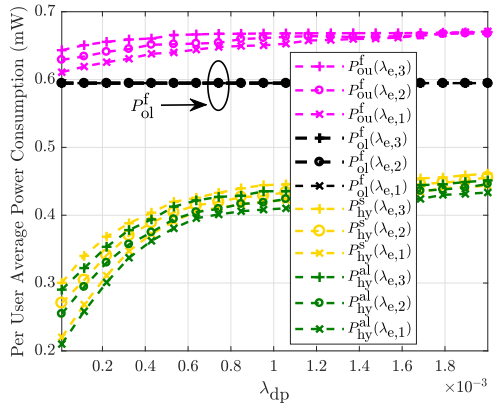}} \caption{Power consumption of the D2D network against the DP intensity}\vspace*{-0.2cm} \label{Fig: power_dp}
	\end{figure}
	
	Firstly, we validate the assumption of $\Phi_{\DHO}$ and $\Phi_{\DHV}$ being homogeneous PPPs. For this, we use DPs and DCs that are spatially distributed according to homogeneous PPPs. Then, the DPs and DCs are randomly paired based on their link lengths. We split them into outband and overlay using a threshold distance $d$. The coverage probability of a typical DC in each band is evaluated using simulations and compared with the theoretical coverage probability obtained by assuming that $\Phi_{\DHO}$ and $\Phi_{\DHV}$ are homogeneous PPPs. Fig. \ref{Fig: modelval} demonstrates that the simulation results closely match our theoretical results, validating the approximation.

	Figure \ref{Fig: inten_ext} illustrates the variation of the D2D link intensity in each band with $\lambda_{\E}$. The subscripts $\textnormal{ou}$ and $\textnormal{ol}$ are used to denote outband and overlay, respectively. The superscripts $\textnormal{s}$, $\textnormal{th}$, $\textnormal{al}$, and $\textnormal{f}$ are used to denote simulation results, theoretical results, iterative optimization and monolithic (fully overlay or underlay) schemes, respectively.  Increasing $\lambda_{\E}$ results in a reduction in D2D links in both outband and overlay networks. The rate of reduction is faster in the outband network. As $\lambda_{\E}$ increases, the threshold distance $d_{\DHO}^{\star}$ is reduced to decrease the user intensity in the outband, such that they do not violate the QoS requirements.  Moreover, this allocates more users to the overlay mode, resulting in higher interference. Therefore, $d_{\DHV}^{\star}$ is also reduced with at a slower rate compared to $d_{\DHO}^{\star}$, to maintain QoS. Furthermore, one can observe that the iterative optimization provides more D2D communication opportunities compared to our approximate solution. However, it requires higher computational time. Therefore, based on the resource and delay constraints of the system, one can choose between the approximate technique and iterative optimization. The theoretical D2D intensities closely follow the results obtained through simulation. Also, it can be seen that D2D opportunities have increased 4-5 times with the hybrid model compared to pure overlay or outband D2D networks, which may result in significant power savings at the infrastructure nodes.

	Fig. \ref{Fig: inten_dp} presents the user intensities in each band with varying DP intensities. At first, increasing $\lambda_{\DP}$ results in a linear increase in user intensities in each band. As $\lambda_{\DP}$ is further increased, the user intensities begin to saturate. The saturation occurs mainly because additional DC-DP pairs cannot be admitted since they will not satisfy the QoS requirements using the D2D mode due to increased interference in each band. Initially, overlay intensity is higher than the outband intensity since the sparse network in low $\lambda_{\DP}$ regime results in a low probability of finding DP-DC pairs with small link lengths to be allocated to outband. Therefore, more D2D links are eligible for the overlay mode. However, as the network becomes more dense, the probability of finding DC-DP pairs with shorter link lengths increases. Therefore, the number of links satisfying the outband threshold will be higher than the number of links satisfying the overlay threshold.
	
	Fig. \ref{Fig: power_dp} compares the average power consumption of a D2D link in the hybrid network, fully overlay network and the fully outband network, under three different $\lambda_{\E}$ values, namely $\lambda_{e,1}=10^{-3}$, $\lambda_{e,2}=1.5\times 10^{-3}$, and $\lambda_{e,3}=2\times 10^{-3}$. As expected, increasing $\lambda_{\E}$ increases the power consumption of the outband networks since higher transmit power is required to maintain the QoS. Also, the power consumption of the fully overlay network is unaffected by $\lambda_{\E}$. One can see that the hybrid network saves nearly $50\%$ of the power compared to the monolithic networks, indicating the energy efficiency of our proposed model. Again, it can be seen that the iterative optimization results in lower power consumption at the devices. However, it may result in higher power consumption at the infrastructure nodes due to the increased complexity.
	
	\section{Conclusion}\label{conclusion}
	A spectrum selection and transmit power minimization scheme was proposed for a D2D network cross-laid with a CRAN, where D2D communications are allowed as both overlay to the CRAN and in the ISM band. Analytical approximations were derived for the spectrum selection thresholds and the required minimum transmit power to achieve a guaranteed QoS level. Theoretical approximations were derived for the D2D user intensity in each band, which can be used to evaluate important performance metrics such as coverage probability and transmission capacity. The proposed scheme achieves nearly 50\% power savings compared to a monolithic D2D network, where D2D communications occur only as overlay or in the ISM band.
	\vspace*{-0.1cm}
	\bibliographystyle{ieeetr}

\end{document}